# Semantic Information Retrieval from Distributed Heterogeneous Data Sources

K. Munir, M. Odeh, R. McClatchey, S. Khan, I. Habib

*CCS Research Centre, University of West of England, Frenchay, Bristol, UK*
*Email :{Kamran.Munir, Richard.Mcclatchey}@cern.ch,*
*Mohammed.Odeh@uwe.ac.uk@uwe.ac.uk, khan@irit.fr, irfan.habib@niit.edu.pk*

**Abstract**

Information retrieval from distributed heterogeneous data sources remains a challenging issue. As the number of data sources increases more intelligent retrieval techniques, focusing on information content and semantics, are required. Currently ontologies are being widely used for managing semantic knowledge, especially in the field of bioinformatics. In this paper we describe an ontology assisted system that allows users to query distributed heterogeneous data sources by hiding details like location, information structure, access pattern and semantic structure of the data. Our goal is to provide an integrated view on biomedical information sources for the Health-e-Child[1] project with the aim to overcome the lack of sufficient semantic-based reformulation techniques for querying distributed data sources. In particular, this paper examines the problem of query reformulation across biomedical data sources, based on merged ontologies and the underlying heterogeneous descriptions of the respective data sources.

## 1. Introduction

Over the past few years, the biomedical domain has been witnessing a tremendous increase in the number of data providers, the volume, and heterogeneity of generated data. To enable knowledge discovery, clinicians' queries generally require an integrated and merged view of the data available across distributed data sources. The associated query processing is therefore based on searching for information in documents, searching within (often very heterogeneous) databases and searching for metadata or descriptions of data. Query reformulation is a part of this query processing whose main objective is to extend a user query in order to retrieve additional meaningful results and to access data from data source(s) according to user needs. In recent years, several methods have been proposed that use semantic knowledge and mapping details to reformulate a user query in order to provide quick and intelligent answers to the queries.

Ontology integration and merging approaches are widely being used for the integration of information from distributed heterogeneous data sources [F. Hakimpour 2001]. In order to effectively utilize an integrated or merged ontology, intelligent query reformulation techniques are often required. The challenging problem here is that query reformulation ought to be based on the merged ontology and the descriptions of underlying heterogeneous data sources with the goal of overcoming the lack of sufficient semantic-based reformulation techniques for querying distributed heterogeneous data sources.

In this position paper we describe a framework for a data integration system which provides access to distributed heterogeneous data sources. Our aim here is to demonstrate how a merged ontology that is constructed over distributed information source ontologies can effectively be exploited to reformulate a user query that suits the needs of the user.

The data integration and semantic information retrieval concept presented in this paper will lead towards the construction of powerful query reformulation rules to be utilized in the European Health-e-Child (HeC) [J. Freund 2006] project. The Health-e-Child project aims to develop an integrated healthcare platform for European paediatrics, providing seamless integration of traditional and emerging sources of biomedical information. The long-term goal of the project is to provide uninhibited access to universal biomedical knowledge repositories for personalised and preventive healthcare, large-scale information-based biomedical research and training, and informed policy making.

In the remaining part of this paper we begin by discussing and analyzing the feasibility of using existing biomedical information integration systems.

---

[1] See www.Health-e-Child.org

We have closely analyzed the two most cited ontology based information integration approaches namely data warehousing and mediation. Finally, after presenting related work in sections 2 and 3 we outline our methodology for semantic data retrieval based on distributed heterogeneous data sources in which we utilise a merged ontology and identify the challenging problem of query reformulation on the basis of merged ontology and data source descriptions.

## 2. Ontologies and the Integration of Biomedical Information Sources.

Today, we are faced with the challenging problems of dealing with distributed and heterogeneous data sources containing huge amounts of data in varieties of semantic structures. Designing a data integration system is a complex task which involves major issues that include the heterogeneity of the underlying data sources, the difference in access mechanisms, the support of query languages and aspects of semantic heterogeneity in relation to their data models. Currently ontologies are being widely used to overcome the problem of semantic heterogeneity. In this paper we introduce an architecture which utilises ontologies for data integration to provide access to distributed heterogeneous data sources. We use a merged ontology and an associated mapping of information which will enable us to construct query reformulation rules for the semantic information retrieval to be utilised in the HeC project.

Currently ontologies are being used as the basis for communication for representing and storing data, for knowledge sharing, classification and organization of data resources and for policy enforcement etc. The term 'ontology' has been defined in many different ways [B. Chandrasekaran 1999, M. Uschold 1996 and C. Wroe 2003]. The simplest definition of ontology is that "it describes the logical structure of a domain, its concepts and the relationships". A domain ontology means an ontology that has been built for a particular subject or for a specific problem in the domain e.g. bioinformatics, geophysics, brain tumors, cardiac disease etc. or for any sub-type of a particular subject e.g. neurons. A number of ontologies have been developed for the purposes of managing and extracting semantic knowledge from on-line literature and databases.

Biomedical information sources are considerably increasing in number and we have an opportunity to speed the progress of biomedical research through computing advances in order to manage and analyze these biomedical data. In order to effectively analyse this data we need to have powerful data integration systems to provide an integrated view of the information. Biomedical information integration systems can be based on a data warehousing or mediation approach which are then enriched by ontologies to manage and extract semantic knowledge. In the following sections, these approaches are discussed in more detail.

### 2.1 The Data Warehousing Approach with an Ontology Based Query Facility

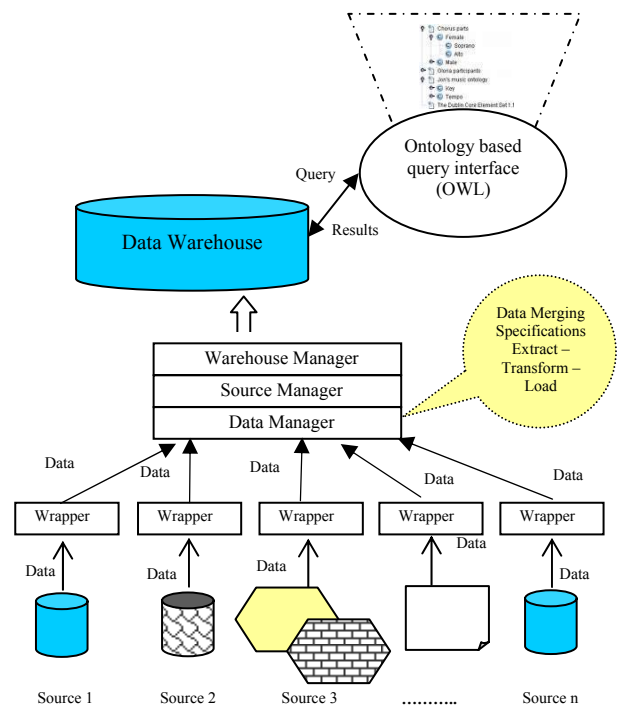

Figure 1: Data warehousing approach with ontology based query facility

A mock-up data integration system utilizing the data warehousing approach by using an ontology based query facility is shown in Figure 1. The data warehousing approach uses a single, centralized data storage to physically retain a copy of the data from each data source. The schema in the data warehouse holds the collective schema of all data sources (called the global schema), and the ontology that is built on top of the global schema is called the global ontology. Here the schema defines the database at the logical level while the ontology defines the database at the conceptual level; mappings are provided between a schema and the ontology to link them. User queries are formulated on the global ontology and all requests are directly answerable by the warehouse. This can results in fast responses and enables multifaceted results from a centralized data store.

However, there are many issues in utilising this kind of system for the integration of biomedical data sources and these include different structural and semantic representations of each data source, data redundancy, data security and data warehouses maintenance. Generally the intention is to avoid duplicating terabytes of data. Moreover, the data available in different biomedical data sources may contain the data about individuals, groups of patients, micro-array data, personal data, team or consortium research data. Each data source may therefore have special arrangements for its storage and access already agreed. Moving all this data from sources into a warehouse involves a huge rebuild of data administration and security infrastructures. Managing a data warehouses is also not a simple task. Whenever new data is added or removed from any of the source systems the update has to be reflected in the warehouse and this may require suspension of the execution of user data requests.

This architecture is often called an information push model, where the data is "pushed" into the data warehouse at definite times. Some of the systems using partially similar approach are also explained in section 3, the literature survey section.

## 2.2 The Mediation Approach with Individual Data Source Ontologies

We have also realized the feasibility of the mediation approach by linking different source systems to wrappers and mediators as a result of their market availability. Many different approaches have been proposed in order to build mediators for information retrieval systems (e.g. see [H. Nottelmann 2001 and N. Fuhr 1999]) and for web-based sources (e.g. see [J. L. Ambite 1998 and C.-C. K. Chang 1999]). A wrapper can be used to make the source available to a mediator and it can handle query processing on individual sources, whereas mediators perform queries on multiple sources. A system utilising the mediation approach with ontologies is shown in Figure 2.

The proposed system for the HeC project utilises ontologies to retrieve data from data sources. In this approach all ontologies are stored in a common repository and when a user submits a query, the system identifies which ontology will be used. Here mediators segregate the user data request into sub-requests with respect to each separate data source by utilising the ontologies with the wrappers providing access to data sources. An overview of a partially similar system called CIRBA was proposed by [Fons Wijnhoven 2003] which employs an ontology-based information retrieval system to solve semantic problems for data market services. This is a form of interactive system where a query is built by asking questions of the user. Utilising this kind of architecture for the integration of a number of biomedical data sources can cause performance overheads where it is required to identify semantic relationship like homonyms and synonyms from all source system ontologies. Secondly each data source is represented by its own ontology and there is no global view of the data. Thirdly there is also an overhead of maintaining the relationships between individual data source ontologies. Finally all of these limitations will also lead towards more complicated results merging.

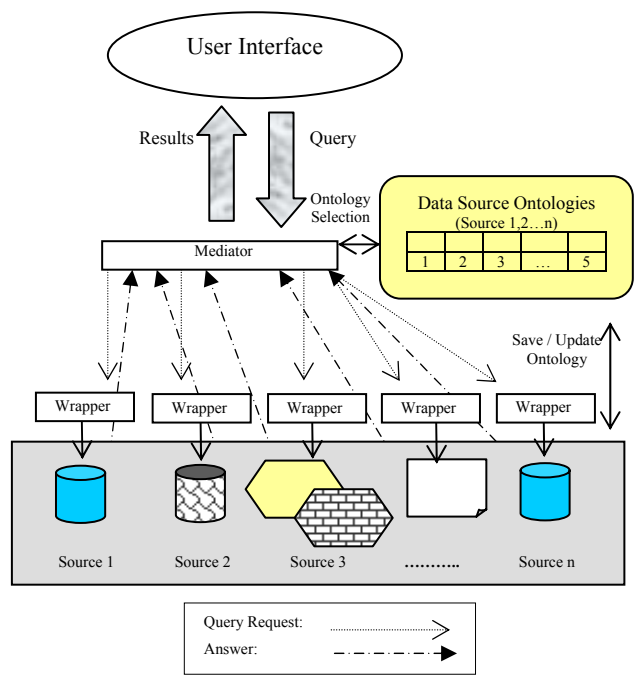

Figure 2: The mediation approach with individual data source ontologies

This kind of architecture is also called an information pull model, where the information is pulled form data sources upon user requests. Some of the other related systems [H. Xiao 2004 and H. Andrade 2004] are explained below in the literature survey section. In [H. Xiao 2004], an approach for the integration and exchange of XML data using RDFS ontologies is presented.

## 3. Related work

There has been previous work carried out at the University of Illinois at Chicago [H. Xiao 2004 & 2006] where they have discussed the problem of data integration and interoperation of heterogeneous XML sources by using an ontology-based framework; they have addressed this problem at the semantic level. They have followed the approach of generating a global ontology by expressing it in RDF (the Resource Description Framework) Schema (RDFS) [D. Brickley 2004]. They discussed a model for mappings between XML schemas and local RDFS ontologies and those between local ontologies and the global RDFS ontology. In the proposed ontology-based approach for the integration of XML sources they have used a Global as View (GAV) approach to model the mappings between the source schemas and the global ontology (called integrated view of the source schemas). They first transformed the heterogeneous XML sources into local RDFS ontologies, which are then merged into the global ontology. In addition to the global ontology their merging process also produces a mapping table, which contains the mapping information between concepts in the global ontology and concepts in the local ontologies.

The approach used in this system cannot be utilized in our HeC architecture due to the fact that XML provides only a surface syntax and cannot handle semantics whereas RDF is a data model for objects and relations and only provides simple semantics for this data model [Tim Berners-Lee 2001]. In HeC we need a more expressive language than RDF as we are integrating data from distributed heterogeneous data sources and the data in these data sources can also contain multimedia data.

In the TAMBIS project [P. G. Baker 1998] a prototype system was developed to provide transparent access across disparate biological databases with concepts specified using descriptionlogic based Ontlogy language, namely DAML+OIL (pls check current language) This conceptual model provides the user with the concepts necessary to construct a wide range of multiple-source queries, and the user interface is used for constructing and manipulating queries. In [N.W. Paton 1999] query processing in the TAMBIS bioinformatics source integration system was presented and it focused in particular on the way source independent concepts in the ontology are related to source-dependent middleware calls, and describing the ways of evaluating user queries. In TAMBIS users are allowed to specify queries transparently over data sources without needing to be aware of the location, capabilities, data types or interfaces of the sources.

However, in TAMBIS the users must be familiar with the content of the ontology and to use a large ontology to navigate and to select appropriate terms makes a number of difficulties. In [N.W. Paton 1999] two other proposals, OBSERVER [E. Mena 2000] and SIMS [Arens 1993], that are close to TAMBIS are reported. In OBSERVER there is no global schema, but rather the emphasis is on peer-to-peer querying among sources, each of which is described using an ontology; in SIMS, the query planner assumes some measure of query processing capabilities from sources. In our architecture we utilise an integrated merged ontology to provide a global view of individual biomedical source ontologies. In this architecture users need not be aware of the content of the underlying individual ontologies to query information sources.

In [C. B. Necib 2003] an approach was presented using ontology knowledge for query processing within a single relational database to extend the result of a query in a semantically meaningful way. They have described how an ontology can be effectively exploited to rewrite a user query into another query such that the new query provides additional meaningful results that satisfy the intention of the user. They have used a query processor to reformulate the query using the ontology associated with that database. The emphasis is that users should not care about where and how the data is organised in the source.

SEMEDA [J.Kohler 2003] can be used to collaboratively edit and maintain ontologies, and to query the integrated databases in real time. However some of the desirable features such as multi-database views and the integration of bioinformatics analysis tools/applications are not available. SEMEDA was developed as a 3-tiered system. It consists of a relational database backend to store ontologies, database metadata and semantic database definitions.

## 4. A Methodology for Semantic Information Retrieval from Distributed Heterogeneous Data Sources

In this section we discuss our approach, shown diagrammatically in figure 3, that resolves the problem of semantic heterogeneity between distributed heterogeneous data sources. The approach presented here is based on the availability/generation of ontologies for each data source and the use of a global merged ontology which defines the integrated and virtual view of the underlying distributed heterogeneous data sources.

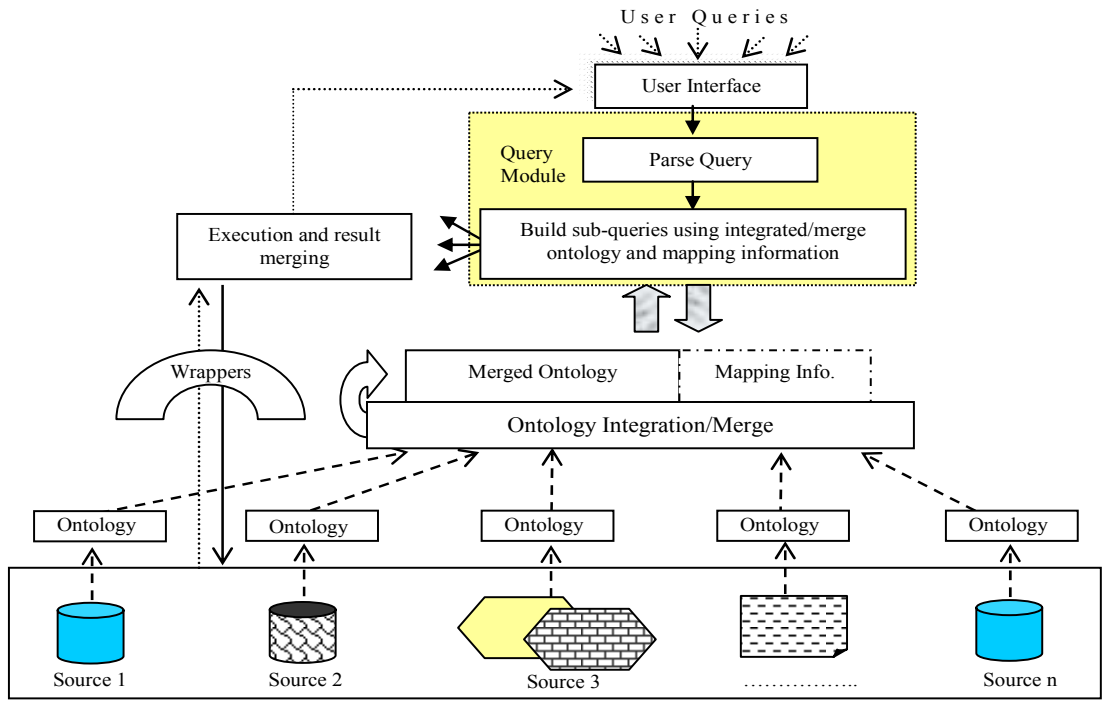

Figure 3: Semantic information retrieval form distributed heterogeneous data sources

The merged ontology provides a unified representation of all underlying ontologies and will be utilised in query generation and reformulation, which can be utilised for knowledge discovery. The reformulation of user queries consists of two steps: data source selection and query translation. When a user query enters the system it is divided into multiple sub-queries for execution at individual data sources; once these queries are executed their results are merged into a final query answer.

At this point in time we are assuming that the structure of the data sources will not change. However, as we are not duplicating or copying source data to any other locations and are using only structural and semantic information of the data to reformulate user queries, data residing in existing arrangements can be deleted and new data can be added at any point in time. Moreover, as explained earlier, while dealing with biomedical data one of the major concerns is the security and confidentiality of data. For each of the data source there can be agreed approved arrangements (subject to suitable ethical clearance) for data storage and access between administrators and users of that particular data source, while utilizing this architecture there is no need to rebuild the whole security infrastructure. Keeping the sources intact will also enable us to keep existing applications running above that data sources.

In order to use this approach there is an overhead in building an ontology for each data source since a detailed merged ontology is required for query resolution. This begs the question:, "Who will build these ontologies?" Ontology development can only be done by the person or by group of people who have a clear understanding of the vocabulary used in the ontology. Ontologies can be reused, extended or partially utilised and considered as long term assets that can be utilized for both resolving semantic conflicts and for communication in different application domains.

A number of pre-existing ontologies are available related to different application domains and there is much research being done on ontology merging and integration. Similarly approaches for ontology merging and integration are also available e.g. [N. F. Noy 2000].The major problem of this architecture is query reformulation where the user's information request is parsed and forwarded for reformulation. In this case, the query is evaluated against a merged ontology and sub-queries are formulated with respect to each of the data source by utilising the mapping information. The problem of reformulation and lack of sufficient semantic-based reformulation techniques for querying distributed heterogeneous data sources is the motivation for our research and is one direction for our future work.

## 5. Conclusion and Future Work

The design of a data integration system can be a complex task and involves major issues to be handled that include: heterogeneity of data, differences in access mechanisms, support for query languages and semantic heterogeneity. In this paper we have described a framework for the data integration system which provides access to distribute heterogeneous data sources by utilising merged ontology and mapping information. Ontologies are used to represent each data source and merged ontology is constructed over individual underlying ontologies to provide a global view. We have also discussed and compared two general data integration approaches that utilises ontologies to provide access to distribute heterogeneous data sources namely data warehouse and mediation approach. Finally we have described the major challenge of reformulation of source independent user query into source specific queries. In future we aim to develop novel approaches using merged ontologies to reformulate a user query into a set of queries that are respectively associated with distributed heterogeneous data source ontologies.

## References

J. Freund, et al., 2006, "Health-e-Child: An Integrated Biomedical Platform for Grid-Based Pediatrics", Studies in Health Technology & Informatics # 120, pp 259-270 IOS Press

F. Hakimpour & A Geppert (2001), "Resolving Semantic Heterogeneity in Schema Integration: An Ontology Based Approach", FOIS'OI, Ogunquit, Maine, USA. ACM l-58113-377-4/01/0010

C. B. Necib & J-C. Freytag, 2003, "Ontology based Query Processing in Database Management Systems", Department of Computer Science Humboldt-Universität zu Berlin.

I. Horrocks, P. F. Patel-Schneider, & F. van Harmelen, 2002, "Reviewing the design of DAML+OIL: an ontology language for the semantic web". In Proc. of the 18th National Conf on Artificial Intelligence, pp. 427--428

C. Wroe, R. Stevens, C. Goble, A. Roberts & M. Greenwood. 2003, A Suite of DAML+OIL Ontologies to Describe Bioinformatics Web Services and Data. Journal of Cooperative Information Science, 198

H. Nottelmann and N. Fuhr, 2001, "MIND: An Architecture for Multimedia Information Retrieval in Federated Digital Libraries". In DELOS Workshop on Interoperability in Digital Libraries, Darmstadt, Germany, September 2001

H. Andrade, T. Kurc et al, 2004, "Optimizing the Execution of Multiple Data Analysis Queries on Parallel and Distributed Environments", IEEE transactions on parallel and distributed systems, Vol. 15, No. 6

H. Xiao & I. F. Cruz,, 2006, "Integrating and exchanging XML data using Ontologies", University of Illinois at Chicago, Journal of Data Semantics

N. Fuhr., 1999, "A Decision-Theoretic Approach to Database Selection in Networkded IR". ACM Transactions on Information Systems, 17(3)

S. Bergamaschi, S. Castano, et. al., 1998, "An intelligent approach to information integration". In Nicola Guarino, editor, Formal Ontologies in Information Systems, pages 253–267. IOS Press.

V. Kashyap & A. Sheth., 1998, "Semantic Heterogeneity in Global Information Systems: the Role of Metadata, Context and Ontologies ". In Cooperative Information Systems: Trends and Directions. Academic Press.

D. Brickley & R. Guha, 2004, RDF Vocabulary Description Language 1.0: RDF Schema. http://www.w3.org/TR/rdf-schema, W3C Working Draft

J. Kohler, S. Philippi & M. Lange, 2003, "SEMEDA: ontology based semantic integration of biological databases", DOI: 10.1093/bioinformatics/btg340, Vol. 19 no. 18, pages 2420–2427

B. Chandrasekaran, J. Josepheson, V.R. Benjamins, 1999, "Ontologies: What are they? Why do we need them?" IEEE Intelligent Systems, 14(1):20-26.

F. Wijnhoven et. al., 2003, "Internal Data Market Services: An Ontology-Based Architecture and Its Evaluation" Informing Science Journal Volume 6

M. Uschold & M. Gruninger, 1999, "Ontologies: principles, methods, and applications", Knowledge Engineering Review, 11(2), 93-155.

J. L. Ambite, N. Ashish et. al., 1998, "A system for constructing mediators for Internet sources". In proc of the ACM SIGMOD International Conference on Management of Data, pages 561–563. (CoopIS'96), Brussels, Belgium, June 1996. IEEE Computer Society Press.

Tim Berners-Lee, James Hendler, Ora Lassila , 2001, "The Semantic Web", Scientific American, Vol. 284, No. 5, pp. 34-43

E. Mena, A. Illarramendi, V. Kashyap & A. Sheth, 2000, "OBSERVER: An approach for query processing in global information systems based on interoperation across pre-existing Ontologies". International journal on Distributed and Parallel Databases (DAPD), 8(2):223–271

Arens, Yigal, et al., 1993, "Retrieving and Integrating Data from Multiple Information Sources". International Journal of Cooperative Information Systems (IJCIS), 2(2):127--158

C.-C. K. Chang & H. Garcia-Molina, 1999. "Mind Your Vocabulary: Query Mapping Across Heterogeneous Information Sources". In Proc. of the ACM SIGMOD, pages 335–346

N. F. Noy M.A. Musen, 2000, "PROMPT: Algorithm and tool for automated ontology merging and alignment". In Proc. of the National Conf on Artificial Intelligence